\documentclass[journal]{IEEEtran}

\usepackage{amsmath,epsfig,bm,amssymb,amsthm}
\usepackage{psfrag}
\usepackage{color}

\usepackage{cite}

\ifCLASSINFOpdf
\else
\fi

\usepackage{amsmath,epsfig,bm,amssymb,amsthm,balance}

\newcommand{\dd}{{\mathrm d}}
\newcommand{\SRi}{{\mathrm{SR}_i}}
\newcommand{\sri}{{\mathrm{sr}_i}}
\newcommand{\RDi}{\mathrm{R}_i\mathrm{D}}
\newcommand{\rdi}{\mathrm{r}_i\mathrm{d}}
\newcommand{\sd}{\mathrm {sd}}
\newcommand{\opt}{\mathrm {opt}}
\newcommand{\Var}{\mathrm {Var}}

\begin{document}
\title{Performance of Differential Amplify-and-Forward Relaying in Multi-Node Wireless Communications
\thanks{Manuscript received November 1, 2012; revised February 21, 2013;
accepted March 20, 2013. Date of publication May 10, 2013. }
\thanks{
This work was supported in part by the Dean's Scholarship from the University of Saskatchewan and a Natural Sciences and Engineering Research Council of Canada Discovery Grant.}
\thanks{
The authors are with the Department of Electrical and Computer Engineering,
University of Saskatchewan, Saskatoon, Canada, S7N5A9.
Email: m.avendi@usask.ca, ha.nguyen@usask.ca.}}

\author{\IEEEauthorblockN{M. R. Avendi,  and Ha H. Nguyen, \textit{Senior Member, IEEE}}
%%\IEEEauthorblockA{Department of Signal and Systems\\
%University of Saskatchewan\\
%Saskatoon, Canada, S7N5A9\\
%Email: mor837@mail.usask.ca, ha.nguyen@usask.ca}
}

\maketitle

\begin{abstract}
\label{abs}
This paper is concerned with the performance of differential amplify-and-forward (D-AF) relaying for multi-node wireless communications over time-varying Rayleigh fading channels. A first-order auto-regressive model is utilized to characterize the time-varying nature of the channels. Based on the second-order statistical properties of the wireless channels, a new set of combining weights is proposed for signal detection at the destination. Expression of pair-wise error probability (PEP) is provided and used to obtain the approximated total average bit error probability (BER). It is shown that the performance of the system is related to the auto-correlation of the direct and cascaded channels and an irreducible error floor exists at high signal-to-noise ratio (SNR). The new weights lead to a better performance when compared to the conventional combining scheme. Computer simulation is carried out in different scenarios to support the analysis.
\end{abstract}

\begin{keywords}
Differential amplify-and-forward relaying, non-coherent detection, time-varying channels, performance analysis, channel auto-correlation, auto-regressive models.

\end{keywords}

\IEEEpeerreviewmaketitle

\section{Introduction}
\label{se:intro}
The increasing demand for better quality and higher data rate in wireless communication systems motivated the use of multiple transmit/receive antennas, resulting in the so-called multiple-input multiple-output (MIMO) systems. However, using multiple antennas is not practical for mobile units due to insufficient space to make wireless channels corresponding to multiple antennas uncorrelated. This limitation was however addressed by the technique of cooperative communications \cite{uplink-Aazgang,user-coop1}, which has been shown to be applicable in many wireless systems and applications such as 3GPP LTE-Advances, WiMAX, WLANs, vehicle-to-vehicle communications and wireless sensor networks \cite{coop-deploy,coop-V2V,coop-WiMAX,coop-dohler,coop-LTE}.

Cooperative communication exploits the fact that, since other users in a network can also listen to a source during the source's transmission phase, they would be able to re-broadcast the received data to the destination in another phase to help the source. Therefore, the overall diversity and performance of the system would benefit from the virtual MIMO system that is constructed using the help of other users. Depending on the strategy that relays utilize to cooperate, the relay networks have been classified as decode-and-forward or amplify-and-forward \cite{coop-laneman}.

Among these strategies, amplify-and-forward (AF) has been the focus of many studies because of its simplicity in the relay's function. Specifically, the relay's function is to multiply the received signal with a fixed or variable gain and forward the result to the receiver. For convenience, the overall channel of source-relay-destination is called the cascaded, the equivalent or double-Rayleigh channel. Depending on the type of modulation, the relays may need full or partial channel state information (CSI) for determining the amplification factor. Also, the destination would need the CSI of both the direct and the cascaded channels in order to combine the received signals for coherent detection.

To avoid channel estimation at the relays and destination, differential AF (D-AF) scheme has been considered in \cite{DAF-Liu,DAF-MN-Himsoon,DAF-DDF-QZ,DAF-General} which only needs the second-order statistics of the channels at the relays. In the absence of instantaneous CSI, a set of fixed weights, based on the second-order statistics, has been used to combine the received signals over the relay-destination and the source-destination links. Then, the standard differential detection is applied to recover the transmitted symbol. However, all the previous works assume a slow-fading situation and show that the performance of D-AF is abound 3-4 dB worse than the performance of its coherent version. For future reference, we call such a scheme ``conventional differential detection'' (CDD). In practice, the increasing speed of mobile users leads to fast time-varying channels (also referred to as time-selective channels). Thus, the typical assumption made in the development of conventional differential detection, namely the approximate equality of two consecutive channel uses, is violated. Therefore, it is important to consider performance of D-AF relaying systems and its robustness under more practical and general channel variation scenarios. It should also be mentioned that the effect of time-varying channels on the performance of coherent AF relay networks has been investigated in \cite{AF-suraweera,AF-Diomidis-2012}.

In this paper, the performance of D-AF for a multi-relay network in \emph{fast} time-varying Rayleigh fading channels is studied. We call the detection scheme developed for such fast time-varying channels ``time-varying differential detection'' (TVD). The channels from the source to the relays (SR channels), the source to the destination (SD channel) and from the relays to the destination (RD channels) are changing continuously according to the Jakes' model \cite{microwave-jake}. Depending on the mobility of nodes with respect to each other, different cases are considered. The direct channel is modelled with a first-order auto-regressive model, AR(1) \cite{AR1-ch,AR2-ch}. Also, based on the AR(1) model of the individual Rayleigh-faded channels, a time-series model is proposed to characterize the time-varying nature of the cascaded channels. The statistical properties of this model are verified using theory and monte-carlo simulation. Taking into account the statistical properties of channel variations, new weights for combining the received signals over multiple channels are proposed. Since analyzing the performance of the proposed system using fixed combining weights is too complicated (if not impossible), the performance of the system using the optimum maximum ratio combining (MRC) weights is analyzed and the result is used as a lower bound for the system error performance. Specifically, the pair-wise error probability (PEP) is obtained and used to approximate the average bit error rate (BER) using nearest neighbour approximation. It is shown that an error floor exits at high signal-to-noise ratio (SNR) region. Such an error floor can be approximately determined and it is related to the auto-correlation values of both the direct and the cascaded channels. Simulation results are presented to support the analysis in various scenarios of fading channels and show that the TVD with the proposed weights always outperforms the CDD in time-selective channels.

The outline of the paper is as follows. Section \ref{sec:system} describes the system model. In Section III the channel model and the differential detection of D-AF relaying with MRC technique over fast time-varying channels is developed. The performance of the system is considered in Section \ref{sec:symbol_error_probability}.
Simulation results are given in Section \ref{sec:sim}. Section \ref{sec:con} concludes the paper.

\emph{Notations:} $(\cdot)^*$, $|\cdot|$ and $\mbox{Re}\{\cdot\}$ denote conjugate, absolute value and the real part of a complex number, respectively. $\mathcal{CN}(0,\sigma^2)$ and $\chi_2^2$ stand for complex Gaussian distribution with mean zero and variance $\sigma^2$ and chi-squared distribution with two degrees of freedom, respectively. $\mbox{E}\{\cdot\}$, $\mbox{Var}\{\cdot\}$ denote expectation and variance operations, respectively. Both ${\mathrm{e}}^{(\cdot)}$ and $\exp(\cdot)$ show the exponential function.

\section{System Model}
\label{sec:system}
\begin{figure}[t]
\psfrag {Source} [] [] [1.0] {Source}
\psfrag {Relay1} [] [] [1.0] {Relay 1}
\psfrag {Relay2} [] [] [1.0] {Relay 2}
\psfrag {RelayR} [] [] [1.0] {Relay R}
\psfrag {Destination} [] [] [1.0] {Destination}
\psfrag {f1} [r] [] [1.0] {$h_{\mathrm{sr}_1}[k]$}
\psfrag {g1} [l] [] [1.0] {$h_{\mathrm{r}_1 \mathrm{d}}[k]$}
\psfrag {f2} [bl] [] [1.0] {$h_{\mathrm{sr}_2}[k]$}
\psfrag {g2} [] [] [1.0] {\;\;$h_{\mathrm{r}_2 \mathrm{d}}[k]$}
\psfrag {fR} [] [] [1.0] {$h_{\mathrm{sr}_R}[k]$\;\;\;}
\psfrag {gR} [l] [] [1.0] {\;\;$h_{\mathrm{r}_R \mathrm{d}}[k]$}
\psfrag {hsd} [] [] [1.0] {\;\;$h_0[k]$}
\centerline{\epsfig{figure={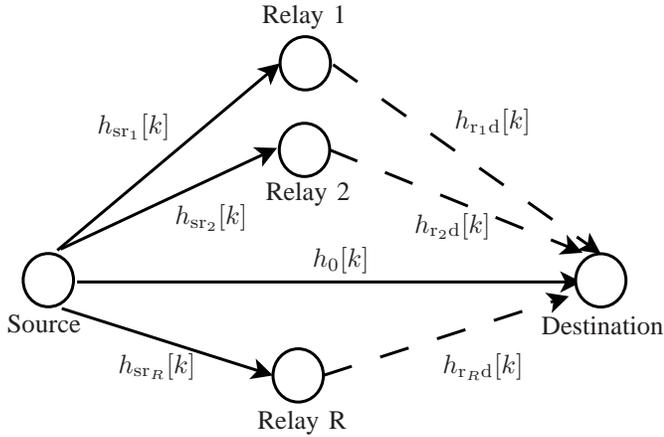},width=8.5cm}}
\caption{The wireless relay model under consideration.}
\label{fig:sysmodel}
\end{figure}

The wireless relay model under consideration is shown in Figure~\ref{fig:sysmodel}. It has one source, $R$ relays and one destination. The source communicates with the destination directly and also via the relays. Each node has a single antenna, and the communication between nodes is half duplex, i.e., each node can either send or receive in any given time. The channels from the source to the destination (SD), from the source to the $i$th relay ($\mathrm{SR}_i$) and from the $i$th relay, $i=1,\cdots,R$, to the destination ($\mathrm{R}_i$D) are shown with $h_0[k]$, $h_{\mathrm{sr}_i}[k]$ and $h_{\mathrm{r}_i\mathrm{d}}[k]$, respectively, where $k$ is the symbol time. A Rayleigh flat-fading model is assumed for each channel. The channels are spatially uncorrelated and changing continuously in time. The auto-correlation value between two channel coefficients, which are $n$ symbols apart, follows the Jakes' fading model \cite{microwave-jake}:
\begin{equation}
\label{eq:jakes-auto}
\mbox{E} \{h[k]h^*[k+n]\}=J_0(2\pi f n),
\end{equation}
where $J_0(\cdot)$ is the zeroth-order Bessel function of the first kind, $f$ is the maximum normalized Doppler frequency of the channel and $h$ is either $h_0$, $h_{\mathrm{sr}_i}$ or $h_{\mathrm{r}_i\mathrm{d}}$. The maximum Doppler frequency of the SD, $\mathrm{SR}_i$ and $\mathrm{R}_i$D channels are shown with $f_{\mathrm{sd}}$, $f_{\mathrm{sr}_i}$ and $f_{\mathrm{r}_i\mathrm{d}}$, respectively. Also, it is assumed that the carrier frequency is the same for all links.

Let $\mathcal{V}=\{{\mathrm{e}}^{j2\pi m/M},\; m=0,\dots, M-1\}$ denote the set of $M$-PSK symbols. At time $k$, a group of  $\log_2M$ information bits is transformed to $v[k]\in \mathcal{V}$. Before transmission, the symbols are encoded differentially as
\begin{equation}
\label{eq:s-source}
s[k]=v[k] s[k-1],\quad s[0]=1.
\end{equation}
The transmission process is divided into two phases. Technically, either symbol-by-symbol or block-by-block dual-phase transmission protocol can be considered. In symbol-by-symbol protocol, first the source sends one symbol to the relays and then the relays re-broadcast the amplified versions of the corresponding received signals to the destination, in a time division manner. Hence, two channel uses are $R+1$ symbols apart. However, this protocol is not practical as it causes frequent switching between reception and transmission. Instead, in block-by-block protocol, a frame of information data is broadcasted in each phase and then two channel uses are one symbol apart. Therefore, block-by-block transmission is considered in this paper. However, the analysis is basically the same for both cases as only the channel auto-correlation values are different.

In phase I, the symbol $\sqrt{P_0}s[k]$ is transmitted by the source to the relays and the destination, where $P_0$ is the average source power. The received signal at the destination and the $i$th relay are
\begin{equation}
\label{eq:source_destination_rx}
y_0[k]=\sqrt{P_0}h_0[k]s[k]+w_0[k]
\end{equation}
\begin{equation}
\label{eq:relay_rx}
y_{\mathrm{sr}_i}[k]=\sqrt{P_0}h_{\mathrm{sr}_i}[k]s[k]+w_{\mathrm{sr}_i}[k]
\end{equation}
where $w_0[k],w_{\mathrm{sr}_i}[k]\sim \mathcal{CN}(0,1)$ are the noise components at the destination and the $i$th relay, respectively.

The received signal at the $i$th relay is then multiplied by an amplification factor $A_i$, and forwarded to the destination. The amplification factor can be either fixed or variable. A variable $A_i$ needs the instantaneous CSI. For D-AF, in the absence of the instantaneous CSI, the variance of the SR channels (here equals to one) is utilized to define the fixed amplification factor as \cite{DAF-Liu,DAF-MN-Himsoon,DAF-DDF-QZ,DAF-General}:
\begin{equation}
\label{eq:A_i}
A_i=\sqrt{\frac{P_i}{P_0+1}}
\end{equation}
where $P_i$ is the average transmitted power of the $i$th relay.

The corresponding received signal at the destination is
\begin{equation}
\label{eq:dest-rx1}
y_i[k]=A_i h_{\mathrm{r}_i\mathrm{d}}[k]y_{\mathrm{sr}_i}[k]+w_{\mathrm{r}_i\mathrm{d}}[k],
\end{equation}
where $w_{\mathrm{r}_i\mathrm{d}}[k]\sim \mathcal{CN}(0,1)$ is the noise component at the destination. Substituting (\ref{eq:relay_rx}) into (\ref{eq:dest-rx1}) yields
\begin{equation}
\label{eq:destination-rx}
y_i[k]= A_i \sqrt{P_0}h_i[k]s[k]+w_i[k],
\end{equation}
where the random variable $h_i[k]=h_{\mathrm{sr}_i}[k]h_{\mathrm{r}_i\mathrm{d}}[k]$ represents the gain of the equivalent double-Rayleigh channel, whose mean and variance equal zero and one, respectively. Furthermore,
$$
%\label{eq:noise at destination}
w_i[k]=A_i h_{\mathrm{r}_i\mathrm{d}}[k]w_{\mathrm{sr}_i}[k]+w_{\mathrm{r}_i\mathrm{d}}[k]
$$
is the equivalent noise component. It should be noted that for a given $h_{\mathrm{r}_i\mathrm{d}}[k]$, $w_i[k]$ and $y_i[k]$ are complex Gaussian random variables with mean zero and variances $\sigma_i^2=A_i^2 |h_{\mathrm{r}_i\mathrm{d}}[k]|^2+1$ and $\sigma_i^2(\rho_i+1)$, respectively, where $\rho_i$ is the average received SNR conditioned on $h_{\mathrm{r}_i\mathrm{d}}[k]$, defined as
\begin{equation}
\label{eq:rhoi}
\rho_i=\frac{A_i^2 P_0 |h_{\mathrm{r}_i\mathrm{d}}[k]|^2}{\sigma_i^2}.
\end{equation}
%Later, in the performance analysis, we take the average over the distribution of $|h_2[k]|^2$.

In the following section, we consider the differential detection of the combined received signals at the destination and evaluate its performance.

\section{Channel Models and Differential Detection}
\label{sec:CDD}
The CDD was developed under the assumption that two consecutive channel uses are approximately equal. However, such an assumption is not valid for fast time-varying channels. To find the performance of differential detection in fast time-varying channels, we need to model both the direct and the cascaded channels with time-series models. Depending on the mobility of the nodes with respect to each other, three cases are considered. The first case applies when a mobile user is communicating with a base station both directly and via other fixed users (or fixed relays) in the network. The second case can happen when the communication between two mobile users are conducted directly and also via other fixed relays. The last case is a situation that a mobile user communicates with another mobile user in the network both directly and with the help of other mobile users. The channel models in these three cases are detailed as follows.

\subsection{Mobile Source, Fixed Relays and Destination}
\label{subsec:mobile-source}
When the source is moving but the relays and the destination are fixed, the SD and all SR channels become time-varying and their statistical properties follow the fixed-to-mobile 2-D isotropic scattering channels \cite{microwave-jake}. However, all RD channels remain static.

First, the direct link is modelled with an AR(1) model \cite{AR1-ch,AR2-ch} as follows
\begin{equation}
\label{eq:AR1-model}
h_0[k]=\alpha_0 h_0[k-1]+\sqrt{1-\alpha_0^2}e_0[k]
\end{equation}
where $\alpha_0=J_0(2\pi f_{\mathrm{sd}}n)\leq 1$ is the auto-correlation of the SD channel and $e_0[k]\sim \mathcal{CN}(0,1)$ is independent of $h_0[k-1]$. Note also that $n=1$ for block-by-block transmission and $n=R+1$ for symbol-by-symbol transmission. The auto-correlation value is equal one for static channels and decreases with higher fade rates. Obviously, this value will be smaller for symbol-by-symbol transmission than for block-by-block transmission, which is another drawback of using symbol-by-symbol transmission in addition to its practical implementation issue.

Similarly, the $\SRi$ channel can be described as
\begin{gather}
\label{eq:AR_hsri}
h_{\mathrm{sr}_i}[k]=\alpha_{\mathrm{sr}_i} h_{\mathrm{sr}_i}[k-1]+\sqrt{1-\alpha_{\mathrm{sr}_i}^2}e_{\mathrm{sr}_i}[k]
\end{gather}
where $\alpha_{\mathrm{sr}_i}=J_0(2\pi f_{\mathrm{sr}_i} n)\leq 1$ is the auto-correlations of the $\mathrm{SR}_i$ channel and $e_{\mathrm{sr}_i}[k]\sim \mathcal{CN}(0,1)$ is independent of $h_{\mathrm{sr}_i}[k-1]$.  Also, under the scenario of fixed relays and destination, two consecutive $\RDi$ channel uses are equal, i.e.,
\begin{equation}
\label{eq:hrdi-fixed}
h_{\mathrm{r}_i\mathrm{d}}[k]=h_{\mathrm{r}_i\mathrm{d}}[k-1].
\end{equation}

Thus, for the $i$th cascaded channel, multiplying \eqref{eq:AR_hsri} by \eqref{eq:hrdi-fixed} gives
\begin{equation}
\label{eq:AR2-model}
h_i[k]=\alpha_{\sri} h_i[k-1] +\sqrt{1-\alpha_{\sri}^2} h_{\mathrm{r}_i\mathrm{d}}[k-1] e_{\mathrm{sr}_i}[k]
\end{equation}
which is an AR(1) model with the parameter $\alpha_{\sri}$ and $h_{\mathrm{r}_i\mathrm{d}}[k-1] e_{\mathrm{sr}_i}[k]$ as the input white noise.

\subsection{Mobile Source and Destination, Fixed Relays}
When both the source and the destination are moving, but the relays are fixed, all the SR and RD channels become time-varying and again follow the fixed-to-mobile scattering model \cite{microwave-jake}. Also, the SD channel follows the mobile-to-mobile channel model \cite{m2m-Akki} which is still Rayleigh fading but with the auto-correlation value of the corresponding model.
Therefore, the AR(1) models in \eqref{eq:AR1-model} and \eqref{eq:AR_hsri} are used for modelling the SD and SR channels, respectively. However, for the SD channel, the value of $\alpha_0$ is obtained from the mobile-to-mobile channel model \cite{m2m-Akki}.

For $\RDi$ channel, the AR(1) model is
\begin{equation}
\label{eq:AR_hrdi}
h_{\mathrm{r}_i\mathrm{d}}[k]=\alpha_{\mathrm{r}_i\mathrm{d}} h_{\mathrm{r}_i\mathrm{d}}[k-1]+\sqrt{1-\alpha_{\mathrm{r}_i\mathrm{d}}^2}e_{\mathrm{r}_i\mathrm{d}}[k]
\end{equation}
where $\alpha_{\mathrm{r}_i\mathrm{d}}=J_0(2\pi f_{\mathrm{r}_i\mathrm{d}} n)\leq1$ is the auto-correlation of $\RDi$ channel and $e_{\mathrm{r}_i\mathrm{d}}[k]\sim \mathcal{CN}(0,1)$ is independent of $h_{\mathrm{r}_i\mathrm{d}}[k-1]$.

Then, for the cascaded channel, multiplying \eqref{eq:AR_hsri} by \eqref{eq:AR_hrdi} gives
\begin{equation}
\label{eq:hsr_hrd}
h_i [k]=\alpha_i  h_i [k-1]+\Delta_i[k],
\end{equation}
where $\alpha_i=\alpha_{\sri}\alpha_{\rdi}\leq 1$ is the equivalent auto-correlation of the cascaded channel and
\begin{multline}
\label{eq:Delta}
\Delta_i[k]  =\alpha_{\mathrm{sr}_i} \sqrt{1-\alpha_{\mathrm{r}_i\mathrm{d}}^2} h_{\mathrm{sr}_i}[k-1] e_{\mathrm{r}_i\mathrm{d}}[k]+\alpha_{\mathrm{r}_i\mathrm{d}} \sqrt{1-\alpha_{\mathrm{sr}_i}^2}\\
h_{\mathrm{r}_i\mathrm{d}}[k-1] e_{\mathrm{sr}_i}[k]
+\sqrt{(1-\alpha_{\mathrm{sr}_i}^2)(1-\alpha_{\mathrm{r}_i\mathrm{d}}^2)}e_{\mathrm{sr}_i}[k]e_{\mathrm{r}_i\mathrm{d}}[k]
\end{multline}
represents the time-varying part of the equivalent channel, which is a combination of three uncorrelated complex-double Gaussian distributions \cite{DGC-M} and uncorrelated to $h_i[k-1]$. Since $\Delta_i[k]$ has a zero mean, its auto-correlation function is computed as
\begin{equation}
\label{eq:var_Delta}
E\{\Delta_i[k]  \Delta_i^*[k+m]\}=
\begin{cases}
1-\alpha_i^2, & \text{if}\;\;  m=0,\\
0, & \text{if} \;\; m\neq 0
\end{cases}
\end{equation}
Therefore, $\Delta_i[k]$ is a white noise process with variance $\mbox{E}\{\Delta_i[k]\Delta_i^*[k]\}=1-\alpha_i^2$.

However, using $\Delta_i[k]$ in the way defined in \eqref{eq:Delta} is not feasible for the performance analysis. Thus, to make the analysis feasible, $\Delta_i[k]$ shall be approximated with an adjusted version of one of its terms as
\begin{equation}
\label{eq:delta_h_hat}
\hat{\Delta}_i[k]= \sqrt{1-\alpha_i^2} {h}_{\mathrm{r}_i\mathrm{d}}[k-1] {e}_{\mathrm{sr}_i}[k]
\end{equation}
which is also a white noise process with first and second order statistical properties identical to that of $\Delta_i[k]$ and uncorrelated to $h_i[k-1]$.

By substituting \eqref{eq:delta_h_hat} into \eqref{eq:hsr_hrd}, the time-series model of the equivalent channel can be described as
\begin{equation}
\label{eq:AR2-model-approx}
h_i[k]=\alpha_i h_i[k-1]+\sqrt{1-\alpha_i^2} h_{\mathrm{r}_i\mathrm{d}}[k-1]e_{\mathrm{sr}_i}[k]
\end{equation}
which is again an AR(1) with parameter $\alpha_i$ and $h_{\mathrm{r}_i\mathrm{d}}[k-1]e_{\mathrm{sr}_i}[k]$ as the input white noise.

Comparing the AR(1) models in \eqref{eq:AR2-model} and \eqref{eq:AR2-model-approx} shows that, in essence, they are only different in the model parameters: the parameter contains the effect of the $\SRi$ channel in the former model, while the effects of both the $\SRi$ and $\RDi$ channels are included in the later model. This means that the model in \eqref{eq:AR2-model-approx} can be used as the time-series model of the cascaded channel for the analysis in both cases. Specifically, for static $\RDi$ channels  $\alpha_{\rdi}=1$ and hence \eqref{eq:AR2-model-approx} turns to \eqref{eq:AR2-model}.

\begin{figure*}[t]
\psfrag {h} [] [] [.8] {$|h_i[k]|$}
\psfrag {delta} [] [] [.8] {\eqref{eq:Delta}}
\psfrag {mdelta} [] [] [.8] {$|\Delta_i[k]|$ or $|\hat{\Delta}_i[k]|$}
\psfrag {deltahat} [] [] [.8] {\eqref{eq:delta_h_hat}}
\psfrag {PDF} [] [] [1] {pdf}
\psfrag {exact} [] [c] [.8] {\eqref{eq:hsr_hrd}}
\psfrag {approx} [] [c] [.8] {\eqref{eq:AR2-model-approx}}
\psfrag {theory} [] [c] [.8] {\eqref{eq:pdf-envelope}}
\centerline{\epsfig{figure={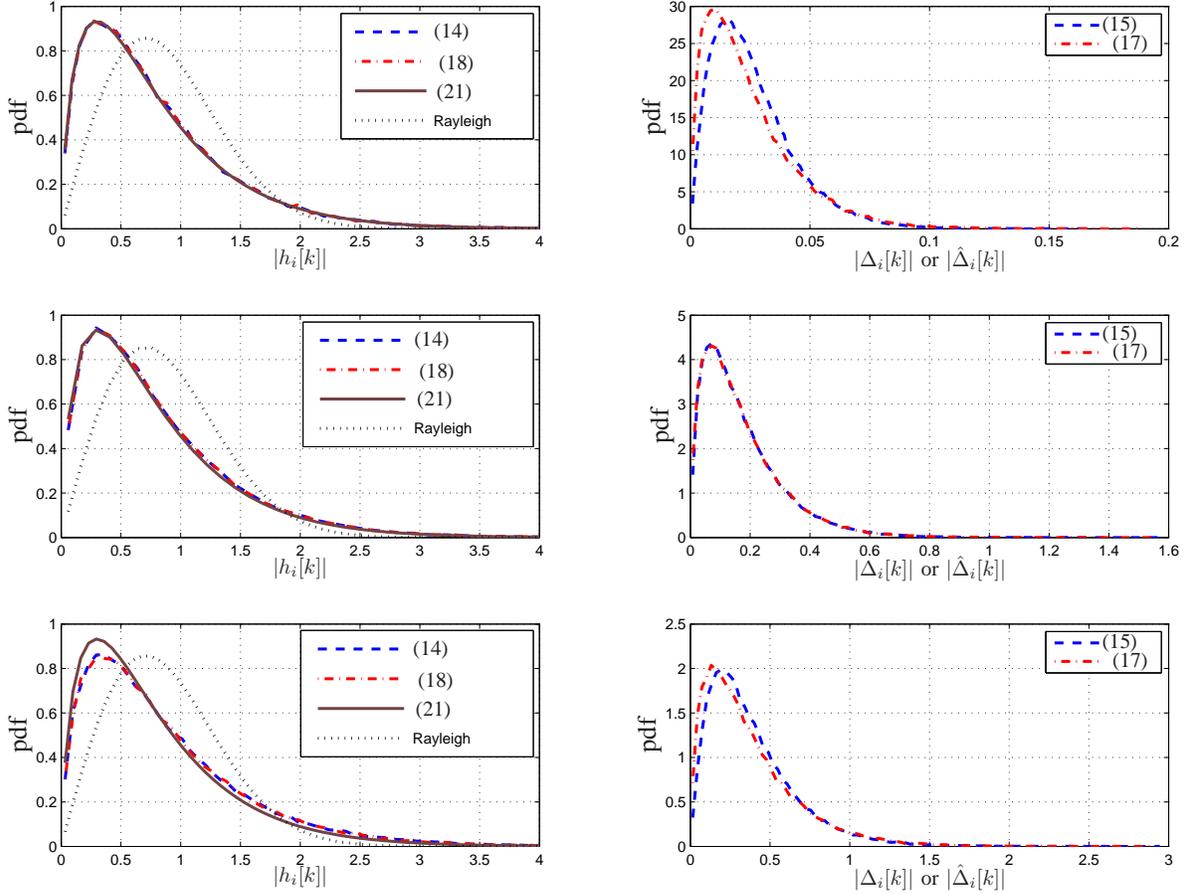},width=19cm}}
\caption{Theoretical pdf of $|h_i[k]|$ and obtained distributions of $|\Delta_i[k]|$, $|\hat{\Delta}_i[k]|$ and $|h_i[k]|$  in Scenario I (upper), Scenario II (middle) and Scenario III (lower). These scenarios are listed in Table \ref{table:scenarios}.}
\label{fig:pdf}
\end{figure*}

To validate the model in \eqref{eq:AR2-model-approx}, its statistical properties are verified with the theoretical counterparts. Theoretical mean and variance of $h_i[k]$ are shown to be equal to zero and one, respectively \cite{SPAF-P,DGC-M}. This can be seen by taking expectation and variance operations over \eqref{eq:AR2-model-approx} so that $\mbox{E}\{h_i[k]\}=0$, $\mbox{Var}\{h_i[k]\}=1$. Also, the theoretical auto-correlation of $h_i[k]$ is obtained as the product of the auto-correlation of the $\SRi$ and $\RDi$ channels in \cite{SPAF-P}. By multiplying both sides of \eqref{eq:AR2-model-approx} with $h_i^*[k-1]$ and taking expectation, one has
\begin{multline}
\label{eq:hi-auto}
\mbox{E}\{h_i[k]h_i^*[k-1]\}=\alpha_i\mbox{E} \{h_i[k-1]h_i^*[k-1]\}\\+\mbox{E}\{\hat{\Delta}_i[k]h_i^*[k-1]\}
\end{multline}
Since $\hat{\Delta}_i[k]$ is uncorrelated to $h_i[k-1]$ then $\mbox{E}\{\hat{\Delta}_i[k]h_i^*[k-1] \}=0$ and it can be seen that
\begin{equation}
\label{eq:h[k]}
\mbox{E}\{h_i[k]h_i^*[k-1]\}=\alpha_i=\alpha_{\mathrm{sr}_i} \alpha_{\mathrm{r}_i\mathrm{d}}.
\end{equation}
In addition, the theoretical pdf of the envelope $\lambda=|h_i[k]|$ is
\begin{equation}
\label{eq:pdf-envelope}
f_{\lambda}(\lambda)=4\lambda K_0\left( 2 \lambda \right)
\end{equation}
where $K_0(\cdot)$ is the zero-order modified Bessel function of the second kind \cite{SPAF-P}, \cite{DGC-M}.
To verify this, using Monte-Carlo simulation the histograms of $|h_i[k]|$, $|\Delta_i[k]|$ and $|\hat{\Delta}_i[k]|$, for different values of $\alpha_i$, are obtained for both models in \eqref{eq:hsr_hrd} and \eqref{eq:AR2-model-approx}. The values of $\alpha_i$ are computed from the normalized Doppler frequencies given in Table \ref{table:scenarios}, which as discussed in Section \ref{sec:sim} covers a variety of practical situations. These histograms along with the theoretical pdf of $|h_i[k]|$ are illustrated in Figure~\ref{fig:pdf}.
Although, theoretically, the distributions of $\Delta_i[k]$ and $\hat{\Delta}_i[k]$ are not exactly the same, we see that for practical values of $\alpha_i$ they are very close. Moreover, the resultant distributions of $h_i[k]$, regardless of $\Delta_i[k]$ or $\hat{\Delta}_i[k]$, are similar and close to the theoretical distribution.
The Rayleigh pdf is depicted in the figure only to show the difference between the distributions of an individual and the cascaded channels.

\subsection{All Nodes are Mobile}
In this case, all links follow the mobile-to-mobile channel model \cite{m2m-Akki}. However, they are all individually Rayleigh faded and the only difference is that the auto-correlation of the channel should be replaced according to this model. Thus, the channel model in \eqref{eq:AR1-model} and \eqref{eq:AR2-model-approx} again can be used as the time-series model of the direct and cascaded channels in this case, albeit with appropriate auto-correlation values. We refer the reader to the discussion in \cite{SPAF-P} and \cite{m2m-Akki} for more details on computing these auto-correlations as well as the tutorial survey on various fading models for mobile-to-mobile cooperative communication systems in \cite{M2M-Talha}. For our analysis, it is assumed that the equivalent maximum Doppler frequency of each link, regardless of fixed-to-mobile or mobile-to-mobile, is given and then the auto-correlation of each link is computed based on \eqref{eq:jakes-auto}.

\subsection{Combining Weights and Differential Detection}
By substituting the time-series models in \eqref{eq:AR1-model} and \eqref{eq:AR2-model-approx} for the direct and the cascaded channels into (\ref{eq:source_destination_rx}) and (\ref{eq:destination-rx}), respectively, one has
\begin{equation}
\label{eq:cddfast-source-destination}
y_0[k]=\alpha_0 v[k] y_0[k-1]+n_0[k],\\
\end{equation}
%where
\begin{equation}
\label{eq:n0}
%\begin{split}
n_0[k]=w_0[k]- \alpha_0 v[k] w_0[k-1]
+ \sqrt{1-\alpha_0^2} \sqrt{P_0} s[k]e_0[k],
%\end{split}
\end{equation}
and
\begin{equation}
\label{eq:cddfast-relay-destination}
y_i[k]=\alpha_i v[k] y_i[k-1]+n_i[k],
\end{equation}
%where
\begin{equation}
\label{eq:ni}
\begin{split}
n_i[k]&=w_i[k]- \alpha_i v[k] w_i[k-1]\\
&+ \sqrt{1-\alpha_i^2}A_i\sqrt{P_0}h_{\mathrm{r}_i\mathrm{d}}[k-1]s[k]e_{\mathrm{sr}_i}[k].
\end{split}
\end{equation}
Note that, the equivalent noise $n_0[k]$ and also $n_i[k]$ for a given $h_{\mathrm{r}_i\mathrm{d}}[k]$, are combinations of complex Gaussian random variables, and hence they are also complex Gaussian with variances
\begin{gather}
\label{eq:variance_eq_nosie1}
\sigma_{n_0}^2=1+\alpha_0^2+(1-\alpha_0^2)P_0\\
\label{eq:variance_eq_nosie2}
\sigma_{n_i}^2=\sigma_i^2 \left(1+\alpha_i^2+(1-\alpha_i^2)\rho_i \right)
\end{gather}

It can be seen that, compared with the CDD scheme, an additional term appears in the noise expression of \eqref{eq:n0} and \eqref{eq:ni} and their variances%%. This term is caused by the channel variation and can be significant in the high SNR region.

To achieve the cooperative diversity, the received signals from the two phases are combined as
\begin{equation}
\label{eq:combined}
\zeta=b_0 y_0^*[k-1]y_0[k]+\sum \limits_{i=1}^R b_i y_i^*[k-1]y_i[k]
\end{equation}
where $b_0$ and $b_i$ are the combining weights. Using the MRC technique \cite{Linear-Diversity}, the optimum combining weights, which takes into account the noise variance of each link, would be
\begin{equation}
\label{eq:optimum_weights}
\begin{split}
b_0^{\mathrm{opt}}&=\frac{\alpha_0}{\sigma_{n_0}^2}\\
b_i^{\mathrm{opt}}&=\frac{\alpha_i}{\sigma_{n_i}^2}, \;\; i=1,\cdots,R
\end{split}
\end{equation}
However, as can be see from \eqref{eq:variance_eq_nosie2}, even for slow-fading channels with $\alpha_i=1$, the noise variance depends on the channel coefficients $h_{\mathrm{r}_i\mathrm{d}}[k]$, which is not known in the system under consideration. To overcome this problem, for slow-fading channels, the average values of the noise variances, $\mbox{E}\{\sigma_{n_0}^2\}=2$ and $\mbox{E}\{\sigma_{n_i}^2\}=2(1+A_i^2)$, were utilized to define the  weights for the CDD scheme as
\begin{equation}
\label{eq:cdd_weights}
\begin{split}
b_0^{\mathrm{cdd}}&=\frac{1}{2}\\
b_i^{\mathrm{cdd}}&=\frac{1}{2(1+A_i^2)}, \;\; i=1,\cdots,R
\end{split}
\end{equation}
It is also shown in \cite{DAF-Liu,DAF-MN-Himsoon,DAF-DDF-QZ,DAF-General} that these weights give a performance close to the optimum combining in slow-fading channels.

For fast time-varying channels, the average variances of the equivalent noise terms in the direct and the cascaded links are $\mbox{E}\{\sigma_{n_0}^2\}=1+\alpha_0^2+(1-\alpha_0^2)P_0$ and $\mbox{E}\{\sigma_{n_i}^2\}=(1+\alpha_i^2)(1+A_i^2)+(1-\alpha_i^2)A^2_i P_0 $, respectively. Therefore, the new combining weights for fast time-varying channels are proposed as
\begin{equation}
\label{eq:b0_b1}
\begin{split}
&b_0=\frac{\alpha_0}{1+\alpha_0^2+(1-\alpha_0^2)P_0}\\
&b_i=\frac{\alpha_i}{(1+\alpha_i^2)(1+A_i^2)+(1-\alpha_i^2)A_i^2P_0}
\end{split}
\end{equation}
It can be seen that for slow-fading, $\alpha_0=1$ and $\alpha_i=1$, which gives $b_0=b_0^{\mathrm{cdd}}$ and $b_i=b_i^{\mathrm{cdd}}$ as expected. However, for fast-fading channels, the weights change with the channel auto-correlation and the source power. In essence, the new weights provide a dynamic combining of the received signals based on the fade rate of each link. The faster the channel changes in a communication link, the smaller portion of the received signal in that link is taken into account for detection. In terms of complexity, the proposed combining weights need the auto-correlation values of the channels which can be computed based on the Jakes' model once the corresponding Doppler frequencies are determined.

Finally, the well known minimum Euclidean distance (ED) detection is expressed as \cite{Dig-ITC-porakis}
\begin{equation}
\label{eq:ml-detection}
\hat{v}[k]= \arg \min \limits_{v[k] \in \mathcal{V}} |\zeta-v[k]|^2
\end{equation}
%which is also equivalent to maximum-likelihood detection with the assumption of given $h_2[k]$ and $n[k]$ Gaussian.
In the next section, we analyse the error performance of this detector.

\section{Error Performance Analysis}
\label{sec:symbol_error_probability}
This section evaluates performance of the D-AF system over time-varying fading channels. Although, the practical combining weights given in \eqref{eq:b0_b1} are used in the detection process, finding the performance of the system with these weights appears infeasible. Instead, performance of the TVD scheme based on the optimum combining weights is carried out and used as a benchmark for the performance of the TVD approach with the proposed weights. It is noted that such an approach in performance analysis is also adopted for the CDD scheme as in \cite{DAF-DDF-QZ,DAF-Liu,DAF-MN-Himsoon}.

Without loss of generality, assume that symbol $v_1$ is transmitted and it is decoded erroneously as $v_2$, the nearest neighbour symbol, by the decoder. The corresponding PEP is defined as $
\label{eq:PEij}
P_s(E_{12})=P_s(v_1\rightarrow v_2).
$
An error occurs when
\begin{equation}
\label{eq:eulidian-distance}
|\zeta-v_1 |^2>|\zeta-v_2|^2
\end{equation}
which can be simplified to
\begin{equation}
\label{eq:pep-cond1}
\text{Re} \left\lbrace (v_1-v_2)^*\zeta  \right\rbrace < 0.
\end{equation}
By substituting $\zeta$ from \eqref{eq:combined} into the above inequality and using $b_0=b_0^{\mathrm{opt}}$ and $b_i=b_i^{\mathrm{opt}}$, the error event can be further simplified as $z>a$ where
\begin{multline}
\label{eq:z_gr_a}
a= |d_{\mathrm{min}}|^2 \left(\alpha_0 b_0^{\mathrm{opt}} |y_0[k-1]|^2+ \sum \limits_{i=1}^{R} \alpha_i b_i^{\mathrm{opt}} |y_i[k-1]|^2\right) \\
z=-2 \text{Re} \bigg\lbrace d_{\mathrm{min}}^* ( b_0^{\mathrm{opt}} y_0^*[k-1] n_0[k] \\
+ \sum \limits_{i=1}^{R} b_i^{\mathrm{opt}} y_i^*[k-1] n_i[k]) \bigg\rbrace
\end{multline}
and $d_{\mathrm{min}}=v_1-v_2$. Note that $n_0[k]$ is Gaussian, while, conditioned on $h_{\mathrm{r}_i\mathrm{d}}[k]$, $n_i[k]$ is also Gaussian. Thus, conditioned on $y_0[k-1]$, $\{y_i[k-1]\}_{i=1}^{R}$ and $\{h_{\mathrm{r}_i\mathrm{d}}[k]\}_{i=1}^R$, the variable $z$ is Gaussian as well. Its mean, $\mu_z$, and variance, $\sigma^2_z$, conditioned on the above variables, are given as (see proof in Appendix \ref{Appen1}):
\begin{multline}
\label{eq:mean-z}
\mu_z=|d_{\mathrm{min}}|^2 \left( \frac{\alpha_0b_0^{\mathtt{opt}}}{P_0+1} |y_0[k-1]|^2 \right.\\
\left. +\sum \limits_{i=1}^{R}\frac{\alpha_i b_i^{\mathrm{opt}}}{\rho_i+1} |y_i[k-1]|^2\right)
\end{multline}

\begin{multline}
\label{eq:var-z}
\sigma_z^2= 2|d_{\mathrm{min}}|^2\left( \alpha_0 b_0^{\mathrm{opt}} |y_0[k-1]|^2\right.\\
\left.+ \sum \limits_{i=1}^{R}\alpha_i b_i^{\mathrm{opt}} |y_i[k-1]|^2 \right)
\end{multline}

Therefore, the conditional PEP can be written as
\begin{multline}
\label{eq:PEP_given_y_h }
P_s(E_{12}|y_0,\{y_i\}_{i=1}^R,\{h_{\mathrm{r}_i\mathrm{d}}\}_{i=1}^R)\\
=\text{Pr}(z> a|y_0,\{y_i\}_{i=1}^R,\{h_{\mathrm{r}_i\mathrm{d}}\}_{i=1}^R) \\
=Q\left( \frac{a-\mu_z}{\sigma_z}\right)
=Q\left(
\sqrt{\Gamma_0+
\sum \limits_{i=1}^{R}\Gamma_i}
\right)
\end{multline}
where $Q(x)=\int \limits_x^{\infty}\frac{1}{\sqrt{2\pi}}\exp\left(\frac{-t^2}{2}\right)\dd t$ and
\begin{gather}
\label{eq:Gamma0}
\Gamma_0=\frac{\gamma_0 |d_{\mathrm{min}}|^2}{P_0+1} |y_0[k-1]|^2\\
\label{eq:Gammai}
\Gamma_i=\frac{\gamma_i |d_{\mathrm{min}}|^2}{\sigma_i^2(\rho_i+1)} |y_i[k-1]|^2
\end{gather}
with
$\gamma_0$ and $\gamma_i$ defined as
\begin{gather}
\label{eq:gamma0}
\gamma_0=\frac{\alpha_0^2 P_0}{2P_0(1-\alpha_0^2)+4+\frac{2}{P_0}}\\
\label{eq:gammai}
\gamma_i=\frac{\alpha_i^2 \rho_i}{2\rho_i(1-\alpha_i^2)+4+\frac{2}{\rho_i}}
\end{gather}

Now, take the average over the distribution of $|y_0[k-1]|^2$ and $|y_i[k-1]|^2$ by using the moment-generating function (MGF) technique \cite{DigComFad-Simon}, the conditional PEP can be written as
\begin{multline}
\label{eq:PEP-h2-MGF}
P_s(E_{12}|\{h_{\mathrm{r}_i\mathrm{d}}\}_{i=1}^R)=\\
\frac{1}{\pi} \int \limits_0^{\pi/2} M_{\Gamma_0} \left( - \frac{1}{2 \sin^2 \theta} \right) \prod \limits_{i=1}^{R} M_{\Gamma_i} \left( - \frac{1}{2 \sin^2 \theta} \right) \dd\theta
\end{multline}
where $M_{\Gamma_0}(\cdot)$ and $M_{\Gamma_i}(\cdot)$ are the MGFs of
$\Gamma_0$ and $\Gamma_i$, respectively.
Since $y_0[k-1]$ and $y_i[k-1]$, conditioned on $h_{\mathrm{r}_i\mathrm{d}}[k]$, are $\mathcal{CN}(0,P_0+1)$ and $\mathcal{CN}(0,\sigma_i^2(\rho_i+1))$, respectively, it follows that $|y_0[k-1]|^2\sim (P_0+1)/2 \chi_2^2$ and $|y_i[k-1]|^2\sim \sigma_i^2(\rho_i+1)/2 \chi_2^2$, respectively.
Hence, the MGFs of $\Gamma_0$ and $\Gamma_i$ can be shown to be \cite{probab-Miller}
\begin{equation}
\label{eq:MGF_Gamma}
\begin{split}
M_{\Gamma_0}(s)=\frac{1}{1-s \gamma_0 |d_{\mathrm{min}}|^2}\\
M_{\Gamma_i}(s)=\frac{1}{1-s \gamma_i |d_{\mathrm{min}}|^2}.
\end{split}
\end{equation}

Therefore, by substituting (\ref{eq:MGF_Gamma}) into (\ref{eq:PEP-h2-MGF}), one obtains
\begin{multline}
\label{eq:PEP-h2-MGF2}
P_s(E_{12}|\{h_{\mathrm{r}_i\mathrm{d}}\}_{i=1}^R)=\\
\frac{1}{\pi} \int \limits_0^{\pi/2} \frac{1}{1+\frac{1}{2\sin^2 \theta}\gamma_0 |d_{\mathrm{min}}|^2} \prod \limits_{i=1}^R\frac{1}{1+\frac{1}{2\sin^2 \theta}\gamma_i |d_{\mathrm{min}}|^2} \dd\theta
\end{multline}
The above integral can be solved by partial fraction technique and then averaged over the distributions of $|h_{\mathrm{r}_i\mathrm{d}}[k]|^2$. However, this leads to a complicated expression without much insight. Instead, we take the average over the distributions of $|h_{\mathrm{r}_i\mathrm{d}}[k]|^2$, $f(\eta_i)=\exp(-\eta_i),\hspace{.1 in} \eta_i>0$, and the unconditioned PEP is given as
\begin{equation}
\label{eq:PEP-integral}
P_s(E_{12})=\\
\frac{1}{\pi} \int \limits_0^{\pi/2} \frac{\prod \limits_{i=1}^R I_i(\theta)}{1+\frac{1}{2\sin^2 \theta}\gamma_0 |d_{\mathrm{min}}|^2}  \dd\theta
\end{equation}
where %$I_i(\theta)$ is given as
\begin{equation}
\label{eq:I1}
\begin{split}
I_i(\theta)&=\int \limits_0^{\infty} \frac{e^{-\eta_i}}{1+\frac{1}{2\sin^2 \theta}\gamma_i |d_{\mathrm{min}}|^2} \dd\eta_i\\
&=\varepsilon_{i}(\theta) \left[ 1+ (\beta_{i}-\epsilon_{i}(\theta)) e^{\epsilon_{i}(\theta)} E_1(\epsilon_{i}(\theta)) \right]
\end{split}
\end{equation}
with $\varepsilon_i(\theta)$, $\beta_i$ and $\epsilon_i(\theta)$ defined as
\begin{multline}
\label{eq: c1_beta1_beta2}
\varepsilon_{i}(\theta)=\frac{4(1-\alpha_i^2)A_i^2P_0+8A_i^2}{\frac{1}{\sin^2(\theta)}\alpha_i^2A_i^2P_0|d_{\mathrm{min}}|^2+4(1-\alpha_i^2)A_i^2P_0+8A_i^2}\\
\beta_i=\frac{4}{2(1-\alpha_i^2)A_i^2P_0+4A_i^2}\\
\epsilon_i(\theta)=\frac{8}{\frac{1}{\sin^2(\theta)}\alpha_i^2A_i^2P_0|d_{\mathrm{min}}|^2+4(1-\alpha_i^2)A_i^2P_0+8A_i^2}
\end{multline}
and $E_1(x)=\int \limits_x^{\infty} ({\mathrm{e}}^{-t}/{t})\dd t$ is the exponential integral function.
The integral in \eqref{eq:PEP-integral}, then can be computed numerically to find the PEP.

It can be verified that, for DBPSK, the expression in \eqref{eq:PEP-integral} gives the exact bit-error rate (BER). On the other hand, for higher-order $M$-PSK constellations, the nearest-neighbour approximation \cite{Dig-ITC-porakis} shall be applied to obtain the overall symbol-error rate (SER) as
$
\label{eq:symbol-error-P}
P_s(E)\approx 2 P_s(E_{12}),
$
and the average BER for Gray-mapping as
\begin{equation}
\label{eq:BER}
P_b(E)\approx \frac{2}{\log_2 M} P_s(E_{12}).
\end{equation}

Finding an upper bound for the PEP expression can help to get more insights about the system performance. For $\theta=\frac{\pi}{2}$, \eqref{eq:PEP-integral} is bounded as
\begin{equation}
\label{eq:upper_bound}
P_s(E_{12})\leq \frac{\prod \limits_{i=1}^RI_i(\frac{\pi}{2})}{2+\gamma_0|d_{\mathrm{min}}|^2}
\end{equation}
Based on the definition of $\gamma_0$ and $I_i({\pi}/{2})$, in \eqref{eq:gamma0} and \eqref{eq:I1}, it can be seen that, the error probability depends on the fading rates, $\alpha_0$ and $\alpha_i$, of both the direct and the cascaded channels. If all channels are very slow-fading, $\alpha_0= 1$ and $\alpha_i= 1$ for $i=1,\ldots, R$, and it can be verified that $\gamma_0 \propto P_0$ and $I_i({\pi}/{2})\propto ({1}/{P_0})$. Thus the diversity order of $R+1$ is achieved. On the other hand, if the channels are fast time-varying, the terms $(1-\alpha_0^2)P_0$ and $(1-\alpha_i^2)P_0$ in the denominator of $\gamma_0$ and $I_i({\pi}/{2})$ become significant in high SNR. This decreases the effective values of $\gamma_0$ and $\gamma_i$ and consequently the overall performance as well as the achieved diversity order of the system will be affected.

It is also informative to examine the expression of PEP at high SNR values. In this case,
\begin{equation}
\label{eq:gama_bar_0}
\bar{\gamma}_0=\lim \limits_{P_0\rightarrow \infty} \gamma_0= \frac{\alpha_0^2}{2(1-\alpha_0^2)}
\end{equation}
and (see proof in Appendix \ref{Appen2})
\begin{equation}
\label{eq:gama_bar_i}
\bar{\gamma}_i=\lim \limits_{P_0\rightarrow \infty} E[\gamma_i]= \frac{\alpha_i^2}{2(1-\alpha_i^2)}
\end{equation}
which is independent of $|h_{\mathrm{r}_i\mathrm{d}}[k]|^2$. Therefore, by substituting the above converged values into \eqref{eq:PEP-h2-MGF2} or \eqref{eq:PEP-integral}, it can be seen that the error floor appears as (see proof in Appendix \ref{Appen3}),
\begin{multline}
\label{eq:PEP-floor}
\lim \limits_{P_0 \rightarrow \infty}P_s(E_{12})=\\
\frac{1}{2}\sum\limits_{k=0}^R
\frac{\bar{\gamma}_k^{R}}{\prod \limits_{\substack{j=0 \\ j\neq k}}^R (\bar{\gamma}_k-\bar{\gamma}_j)}
\left\lbrace 1-\sqrt{\frac{\bar{\gamma}_k |d_{\mathrm{min}}|^2}{2+\bar{\gamma}_k |d_{\mathrm{min}}|^2}}
\right\rbrace
\end{multline}
when $\bar{\gamma}_k\neq \bar{\gamma}_j, \forall \; k,j \geq 0$

\begin{multline}
\label{eq:PEP-floor2}
\lim \limits_{P_0 \rightarrow \infty}P_s(E_{12})=\\
\frac{1}{2} \left\lbrace
1- \sqrt{\frac{\bar{\gamma}|d_{\mathrm{min}}|^2}{\bar{\gamma}|d_{\mathrm{min}}|^2+2}} \sum \limits_{l=0}^{R} \binom{2l}{l} \left( \frac{1}{4+2\bar{\gamma}|d_{\mathrm{min}}|^2} \right)^l
\right\rbrace
\end{multline}
when $\; \bar{\gamma}_0=\bar{\gamma}_i=\bar{\gamma}, \forall \; i>0$

\begin{multline}
\label{eq:PEP-floor3}
\lim \limits_{P_0 \rightarrow \infty}P_s(E_{12})=\\
\frac{\bar{\gamma}_0^R}{2(\bar{\gamma}_0-\bar{\gamma})^R}\left\lbrace 1- \sqrt{\frac{\bar{\gamma}_0|d_{\mathrm{min}}|^2}{\bar{\gamma}_0|d_{\mathrm{min}}|^2+2}} \right\rbrace-
\sum \limits_{k=1}^{R} \frac{\bar{\gamma}_0^{R-k}\bar{\gamma}}{2(\bar{\gamma}_0-\bar{\gamma})^{R-k+1}}\\
\left\lbrace
1- \sqrt{\frac{\bar{\gamma} |d_{\mathrm{min}}|^2}{\bar{\gamma} |d_{\mathrm{min}}|^2+2}} \sum \limits_{l=0}^{k-1} \binom{2l}{l} \left( \frac{1}{4+2\bar{\gamma}|d_{\mathrm{min}}|^2}\right)^l
\right\rbrace\\
\end{multline}
when $\; \bar{\gamma}_0\neq \bar{\gamma}_i=\bar{\gamma}, \forall \; i>0$

%Another case is when the nodes are divided into several groups such that the nodes in each group have a similar converged value but different from other groups. The expression of this case is ignored here because of its complexity, although it could be obtained using partial fraction technique.

It should be noted that the PEP and the error floor expressions are obtained based on the optimum combining weights and hence, as will be observed in the simulation results, they give a lower bound for the PEP and error floor of the system using the proposed weights. The superior performance of the proposed TVD scheme over the CDD scheme as illustrated in the next section comes with the price of requiring the channel auto-correlations for determining the new combining weights. The accurate determination of these auto-correlations is important since it would affect both the actual system performance and the performance analysis.

\section{Simulation Results}
\label{sec:sim}
In this section a typical multi-node D-AF relay network is simulated in different channel scenarios and for the case that all nodes are mobile (the general case). In all simulations, the channels $h_0[k]$, $\{h_{\mathrm{sr}_i}[k]\}_{i=1}^{R}$ and $\{h_{\mathrm{r}_i \mathrm{d}}[k]\}_{i=1}^{R}$ are generated individually according to the simulation method of \cite{ch-sim}. Based on the normalized Doppler frequencies of the channels, three different scenarios are considered: (I) all the channels are fairly slow fading, (II) the SD and SR channels are fairly fast, while the RD channels are fairly slow, (III) the SD and SR channels are very fast and the RD channels are fairly-fast fading. The normalized Doppler frequencies of the three scenarios are shown in Table \ref{table:scenarios}. The values in the table can be translated to different vehicle speeds of communication nodes in typical wireless systems. For example, in a system with carrier frequency $f_c=2$ GHz and symbol duration $T_s=0.1$ ms, the corresponding Doppler shifts for the SD channel would be around $f_D={f_{\sd}}/{T_s}=50,\; 500,\; 1000$ Hz, which would correspond to the speeds of $v={cf_D}/{f_c}=25,\; 270,\; 540$ km/hr, respectively, where $c=3\times 10^8$ m/s is the speed of light. Usually, the value of 75 km/hr is assumed for a typical vehicle speed in the literature but much faster speeds are common in vehicles such as hi-speed trains. Thus, Table \ref{table:scenarios} covers a wide range of practical situations, from very slow to very fast fading, and these situations can be applicable in present and future wireless applications. In fact, Scenario I is practically equivalent to the case of static channels.
%\vspace*{-0.25cm}
\begin{table}[!h]
\begin{center}
\caption{Three simulation scenarios.}
\label{table:scenarios}
  \begin{tabular}{ |c | c| c| c | }
    \hline
				& $f_{\mathrm{sd}}$ & $f_{\mathrm{sr}_i}$ & $f_{\mathrm{r}_i\mathrm{d}}$  \\ \hline\hline
{Scenario I}    & .005              & .005              & .005  			 \\ \hline
{Scenario II}   & .05               & .05 			    & .005   			  \\ \hline
{Scenario III}  & .1               & .1 			    & .05   			  \\
    \hline
  \end{tabular}
\end{center}
\end{table}
%\vspace*{-0.5cm}

\begin{figure}[tb]
\psfrag {P(dB)} [][] [.8]{$P$ (dB)}
\psfrag {BER} [] [] [.8] {BER}
\psfrag {Scenario I} [] [] [.8] {Scenario I}
\psfrag {Scenario II} [] [] [.8] {Scenario II}
\psfrag {Scenario III} [b] [] [.8] {Scenario III}
%\psfrag {Slow-Fading} [] [cr] [.8] {Slow-fading}
\psfrag {Analysis} [l] [l] [.75] {\small Lower Bound}
\centerline{\epsfig{figure={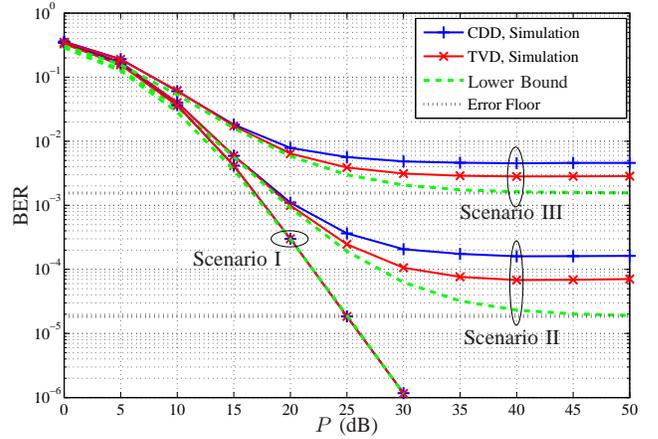},width=8.5cm}}
\caption{Theoretical and simulation results of D-AF relaying with two relays in Scenario I (lower plots), Scenario II (middle plots) and Scenario III (upper plots) using DBPSK.}
\label{fig:r2_m2_scs}
\end{figure}

\begin{figure}[tb]
\psfrag {P(dB)} [][] [.8]{$P$ (dB)}
\psfrag {BER} [] [] [.8] {BER}
\psfrag {Scenario I} [] [] [.8] {Scenario I}
\psfrag {Scenario II} [] [] [.8] {Scenario II}
\psfrag {Scenario III} [] [] [.8] {Scenario III}
\psfrag {Analysis} [l] [l] [.75] {\small Lower Bound }
\centerline{\epsfig{figure={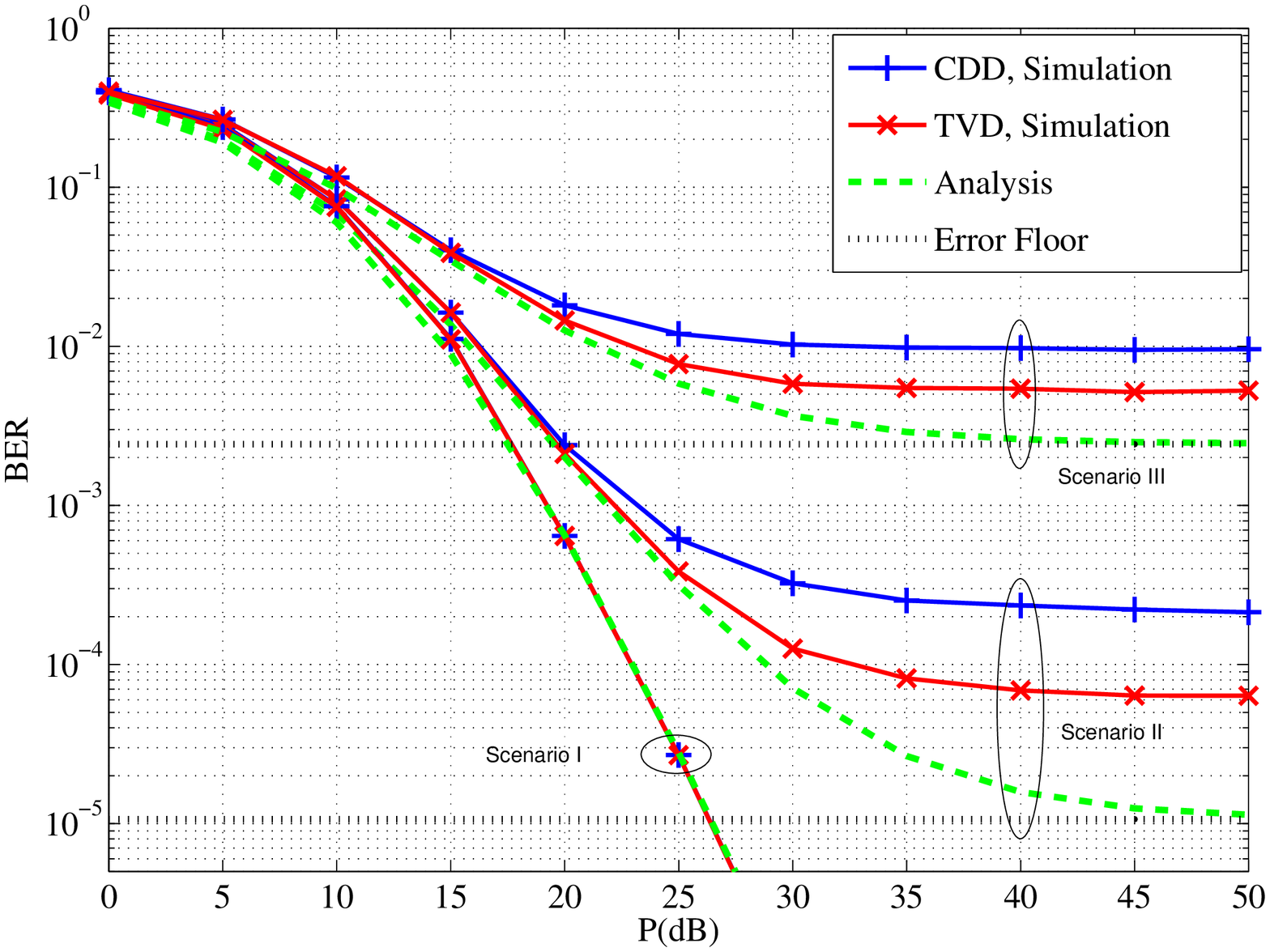},width=8.5cm}}
\caption{ Theoretical and simulation results of D-AF relaying with three relays in Scenario I (lower plots), Scenario II (middle plots) and Scenario III (upper plots) using DQPSK.}
\label{fig:r3_m4_scs}
\end{figure}

In each scenario, binary data is differentially encoded for DBPSK ($M=2$) or DQPSK ($M=4$) constellations. Block-by-block transmission is conducted in all scenarios. The amplification factor at the relay is fixed to $A_i=\sqrt{{P_i}/{(P_0+1)}}$ to normalize the average relay power to $P_i$. The power allocation among the source and relays is such that $P_0={P}/{2}$ and $P_i={P}/{(2R)}$, where $P$ is the total power consumed in the network. Note that, due to the way the variance of all AWGN components and channel gains is normalized to unity, the total power $P$ also has the meaning of a signal-to-noise ratio (SNR). At the destination, the received signals are first combined with the proposed weights so that the minimum Euclidean-distance detection can then be carried out. The simulation is run for various values of the total power in the network. For comparison, the same simulation process but with the combining weights given in \eqref{eq:cdd_weights} is repeated for the CDD system. The practical BER values obtained with the CDD and TVD schemes are plotted versus $P$ in Fig.~\ref{fig:r2_m2_scs} (solid lines but different markers) for DBPSK and a two-relay network. Fig. \ref{fig:r3_m4_scs} shows similar BER plots but for DQPSK and a three-relay network.

On the other hand, for computing the theoretical BER values, first the values of $\alpha_i$ and $\alpha_0$ are computed for each scenario. Also, $|d_{\min}|^2=4   \sin^2({\pi}/{M})$ for $M$-PSK symbols is computed to give $|d_{\min}|^2=4$ for $M=2$, and $|d_{\min}|^2=2$ for $M=4$. Then, the corresponding theoretical BER values from \eqref{eq:BER} are plotted in the two figures with dashed lines.

As can be seen from Figs.~\ref{fig:r2_m2_scs} and ~\ref{fig:r3_m4_scs}, in Scenario I of very slow-fading (practically the scenario of static channels), the desired cooperative diversity is achieved with both the CDD and TVD schemes. The BER curves for both schemes monotonically decrease with increasing $P$ and are consistent with the theoretical values. Since in this scenario, all the channels are fairly slow, the combining weights are approximately equal in both CDD and TVD systems and the BER results are very tight to the theoretical values which are determined using the optimum combining weights. Also, the error floor is very low and does not practically exist in this slow-fading situation and it is not plotted.

In Scenario II, which involves two fast-fading channels, the BER plots gradually deviate from the BER results obtained in Scenario I, at around 15 dB, and reach an error floor for $P\geq 30$ dB. The error floor is also calculated theoretically from (\ref{eq:PEP-floor3}) and plotted in the figures with dotted lines. The error floor is around $6\times 10^{-5}$ for TVD scheme, while it is around $2\times 10^{-4}$ for the CDD scheme in both figures. The significantly-lower error floor of the TVD scheme clearly shows its performance improvement over the CDD scheme.  The ``deviating'' phenomenon starts earlier, around 10 dB in Scenario III, and the performance degradation is much more severe since all the channels are fast fading in this scenario. Although the existence of the error floor is inevitable in both detection approaches, the TVD scheme with the proposed weights always outperforms the CDD scheme and it performs closer to the theoretical results using the optimum weights. %Generally speaking, for a medium value of $P$, around 5 dB can be saved with the TVD scheme as compared to the CDD scheme in both Scenarios II and III.
As expected, for both Scenarios II and III, the theoretical BER plots corresponding to the optimum combining weights give lower bounds for the actual performance. Another important observation is that the achieved diversity is severely affected by the high fade rates of time-varying fading channels, although the multiple fading channels are still independent. %The information about the error floor would help the system designer to prevent the system from operating in such an error floor region. One possibility is to mitigate the effect of channel variations by deploying more relays in the network.

%\begin{figure}[tb]
%\psfrag {P(dB)} [][] [.8]{$P$ (dB)}
%\psfrag {BER} [] [] [.8] {BER}
%%\psfrag {Scenario I} [] [cr] [.8] {Scenario I}
%%\psfrag {Scenario II} [] [cr] [.8] {Scenario II}
%%\psfrag {Scenario III} [] [cr] [.8] {Scenario III}
%\psfrag {Theoritical Optimum} [] [c] [.6] {Theoretical Optimum}
%\centerline{\epsfig{figure={figure/r2_m4_scs.eps},width=8.5cm}}
%\caption{ Theoretical and simulation results of D-AF relaying with two relays in Scenario I (lower plots), Scenario II (middle plots) and Scenario III (upper plots) using DQPSK.}
%\label{fig:r2_m4_scs}
%\end{figure}

%\begin{figure}[tb]
%\psfrag {P(dB)} [][] [.8]{$P$ (dB)}
%\psfrag {BER} [] [] [.8] {BER}
%%\psfrag {Scenario I} [] [cr] [.8] {Scenario I}
%%\psfrag {Scenario II} [] [cr] [.8] {Scenario II}
%%\psfrag {Scenario III} [] [cr] [.8] {Scenario III}
%%\psfrag {Theoritical Optimum} [] [c] [.6] {Theoretical Optimum}
%\centerline{\epsfig{figure={figure/r3_m2_scs.eps},width=8.5cm}}
%\caption{ Theoretical and simulation results of D-AF relaying with three relays in Scenario I (lower plots), Scenario II (middle plots) and Scenario III (upper plots) using DBPSK.}
%\label{fig:r3_m2_scs}
%\end{figure}

%In another experiment we compare the symbol-by-symbol and block-by-block transmission protocols. The simulation is run for the two relay network in Scenario II and the average BER for both transmissions are plotted in Figure~\ref{fig:r2_m2_sc2_sym_vs_blc}. As we mentioned before, in symbol-by-symbol transmission, two channel uses are $R+1$ symbols apart in compare with one symbol apart for block-by-block transmission. Hence, the overall auto-correlations are smaller for symbol-by-symbol transmission at similar channel fade rate, which leads to higher degradation in the performance. This effect can be clearly seen in the figure. In the figure, the upper plots correspond to symbol-by-symbol transmission and the lower plots correspond to block-by-block transmission protocol. %As we can see, an average BER of $10^{-2}$ is the best performance that we can achieve for symbol-by-symbol transmission in compare with $6 \times 10^{-5}$ in block-by-block transmission.

%Now, we repeat the above experiment for a three relays network in Scenario II. The results can be seen in Figure~\ref{fig:r3_m2_sc2_sym_vs_blc}. Despite our expectation, adding one relay to the network is more disruptive than helpful for the performance of symbol-by-symbol transmission. This comes from the fact that the two channel uses become more apart and hence the overall auto-correlations reduce. The experiments suggest that using symbol-by-symbol transmission should be carefully considered in D-AF relay networks, specially when channels are time-selective.

\section{Conclusion}
\label{sec:con}
Performance of multi-node relay networks has been analyzed when differential $M$-PSK modulation along with the amplify-and-forward strategy are used over fast time-varying channels. The time-varying nature of the channels was related to their auto-correlation values. Using the auto-correlation values, the new combining weights at the destination were provided. The obtained error probability expression serves as a lower bound of the actual BER. It was shown that the error performance depends on the fading rates of the direct and the cascaded channels. For fast fading channels, a large fading rate can lead to a severe degradation in the error probability. It was also shown that there exists an error floor at high SNR in time-varying channels and such an error floor was determined in terms of the channel auto-correlations. The analysis is supported with simulation in different scenarios and depicts that the proposed combining gains lead to a better performance over that achieved with the conventional combining weights.

%\begin{figure}[b]
%\psfrag {P(dB)} [][] [.8]{$P$ (dB)}
%\psfrag {BER} [] [] [.8] {BER}
%\centerline{\epsfig{figure={figure/r2_m2_sc2_sym_vs_blc.eps},width=8.5cm}}
%\caption{ Comparing symbol-by-symbol (upper plots) and block-by-block (lower plots) transmission protocols for two relays network using DBPSK}
%\label{fig:r2_m2_sc2_sym_vs_blc}
%\end{figure}
%
%\begin{figure}[b]
%\psfrag {P(dB)} [][] [.8]{$P$ (dB)}
%\psfrag {BER} [] [] [.8] {BER}
%\centerline{\epsfig{figure={figure/r3_m2_sc2_sym_vs_blc.eps},width=8.5cm}}
%\caption{ Comparing symbol-by-symbol (upper plots) and block-by-block (lower plots) transmission protocols for three relays network using DBPSK}
%\label{fig:r3_m2_sc2_sym_vs_blc}
%\end{figure}

\appendices

\section{PROOF OF \eqref{eq:mean-z} and \eqref{eq:var-z}}\label{Appen1}
\begin{multline}
\label{eq:proof of mu_z}
\mu_z=\mbox{E}\{z|y_0[k-1],\{y_i[k-1]\}_{i=1}^R,\{h_{\rdi}[k-1]\}_{i=1}^R\}=\\
-2 \text{Re} \left\lbrace d_{\mathrm{min}}^* ( b_0^{\mathrm{opt}} y_0^*[k-1] \mbox{E}\{n_0[k]|y_0[k-1]\} \right.\\ \left.
+ \sum \limits_{i=1}^{R} b_i^{\mathrm{opt}} y_i^*[k-1] \mbox{E}\{n_i[k]|y_i[k-1],h_{\rdi}[k-1]\} ) \right\rbrace\\
=2\alpha_0 \text{Re} \left\lbrace d_{\mathrm{min}}^* ( b_0^{\mathrm{opt}} y_0^*[k-1] \mbox{E}\{w_0[k-1]|y_0[k-1]\} \right.\\ \left.
+ \sum \limits_{i=1}^{R} b_i^{\mathrm{opt}} y_i^*[k-1] \mbox{E}\{w_i[k-1]|y_i[k-1],h_{\rdi}[k-1]\} ) \right\rbrace
\end{multline}
The conditional means of Gaussian noise components $w_0[k-1]$ and $w_i[k-1]$ are obtained as \cite{prob-papo}
\begin{gather}
\label{eq:Ew0-y0}
\mbox{E}\{w_0[k-1]|y_0[k-1]\}=\frac{1}{P_0+1} d_{\min} y_0[k-1]\\
\label{eq:Ewi-yi}
\mbox{E}\{w_i[k-1]|y_i[k-1],h_{\rdi}[k-1]\}=\frac{1}{\rho_i+1} d_{\min} y_i[k-1]
\end{gather}
Substituting \eqref{eq:Ew0-y0} and \eqref{eq:Ewi-yi} into \eqref{eq:proof of mu_z} gives \eqref{eq:mean-z}.

\begin{multline}
\label{eq:proof of var_z}
\sigma^2_z=\Var \{z|y_0[k-1],\{y_i[k-1]\}_{i=1}^R,\{h_{\rdi}[k-1]\}_{i=1}^R\}\\
=2  |d_{\min}|^2 \left( \left(b_0^{\mathrm{opt}}\right)^2 |y_0[k-1]|^2 \mbox{Var}\{n_0[k]|y_0[k-1]\} \right. + \\  \left.
\sum \limits_{i=1}^{R} \left(b_i^{\mathrm{opt}}\right)^2 |y_i[k-1]|^2 \mbox{Var}\{n_i[k]|y_i[k-1],h_{\rdi}[k-1]\} \right)
\end{multline}
The conditional variances of $n_0[k-1]$ and $n_i[k-1]$ are obtained as
\begin{equation}
\label{eq:Var-n0-y0}
\mbox{Var}\{n_0[k-1]|y_0[k-1]\}=1+\alpha_0^2+(1-\alpha_0^2)P_0=\frac{\alpha_0}{b_{0}^{\opt}}
\end{equation}
\vspace{-.2in}
\begin{multline}
\label{eq:Var-ni-yi}
\mbox{Var}\{n_i[k-1]|y_i[k-1],h_{\rdi}[k-1]\}=\\1+\alpha_i^2+(1-\alpha_i^2)\rho_i
=\frac{\alpha_i}{b_{i}^{\opt}}
\end{multline}
Substituting \eqref{eq:Var-n0-y0} and \eqref{eq:Var-ni-yi} into \eqref{eq:proof of var_z} gives \eqref{eq:var-z}.
It should be noted that since $z$ is proportional to the real part of $n_0[k]$ and $n_i[k]$, its variance is proportional to half of the total variance.

\section{PROOF OF \eqref{eq:gama_bar_i}}\label{Appen2}

By substituting \eqref{eq:rhoi} into \eqref{eq:gammai} we have
\begin{equation}
\begin{split}
\lim \limits_{P_0\rightarrow \infty} \mbox{E}\{\gamma_i\}&= \mbox{E} \{ \lim \limits_{P_0 \rightarrow \infty} \gamma_i \}= \\
& \mbox{E} \left\lbrace
\lim \limits_{P_0\rightarrow \infty}
\frac{\alpha_i^2 A_i^2 P_0 \eta_i}
{
\left(
2A_i^2P_0(1-\alpha_i^2) +4A_i^2
\right)\eta_i+4
}
\right\rbrace \\
&= \mbox{E} \left\lbrace
\frac{\alpha_i^2}{2(1-\alpha_i^2)}
\right\rbrace =\frac{\alpha_i^2}{2(1-\alpha_i^2)}
\end{split}
\end{equation}

\section{PROOF OF \eqref{eq:PEP-floor}-\eqref{eq:PEP-floor3}}\label{Appen3}

\begin{multline}
%\begin{split}
\lim \limits_{P_0 \rightarrow \infty} P_s(E_{12})=
\lim \limits_{P_0 \rightarrow \infty} \frac{1}{\pi} \int \limits_{0}^{\frac{\pi}{2}}
\frac{\prod \limits_{i=1}^{R} I_i(\theta)}{1+\frac{1}{2sin^2(\theta)}\gamma_0|d_{\mathrm{min}}|^2} \dd \theta=\\
 \frac{1}{\pi} \int \limits_{0}^{\frac{\pi}{2}}
\frac{\lim \limits_{P_0 \rightarrow \infty}\prod \limits_{i=1}^{R} I_i(\theta)}{\lim \limits_{P_0 \rightarrow \infty}\left( 1+\frac{1}{2sin^2(\theta)}\gamma_0|d_{\mathrm{min}}|^2 \right)} \dd \theta=\\
\frac{1}{\pi} \int \limits_{0}^{\frac{\pi}{2}}
\frac{1}{ 1+\frac{1}{2\sin^2(\theta)}\bar{\gamma}_0|d_{\mathrm{min}}|^2 }
\prod \limits_{i=1}^{R} \frac{1}{ 1+\frac{1}{2\sin^2(\theta)}\bar{\gamma}_i|d_{\mathrm{min}}|^2 }
\dd \theta\\
%\end{split}
\end{multline}

Now, for the first case that $\bar{\gamma}_k \neq \bar{\gamma}_j, \forall \; k,j\geq 0$, using the partial fraction technique gives
\begin{multline}
\label{eq:partial-fraction}
\frac{1}{ 1+\frac{1}{2\sin^2(\theta)}\bar{\gamma}_0|d_{\mathrm{min}}|^2 }
\prod \limits_{i=1}^{R} \frac{1}{ 1+\frac{1}{2\sin^2(\theta)}\bar{\gamma}_i|d_{\mathrm{min}}|^2 }=\\
\prod \limits_{k=0}^{R} \frac{1}{ 1+\frac{1}{2\sin^2(\theta)}\bar{\gamma}_k|d_{\mathrm{min}}|^2 }=\\
\sum \limits_{k=0}^{R} c_k \bar{\gamma}_k \frac{1}{1+\frac{1}{2\sin^2(\theta)} \bar{\gamma}_k|d_{\min}|^2}
\end{multline}
where
$c_k=\frac{\bar{\gamma}_k^{R-1}}{\prod \limits_{\substack{j=0 \\j\neq k}}^{R}(\bar{\gamma}_k-\bar{\gamma}_j)}$.
Then,
\begin{multline}
\label{eq:err-floor-proof1}
\frac{1}{\pi} \int \limits_{0}^{\frac{\pi}{2}}
\sum \limits_{k=0}^{R} c_k \bar{\gamma}_k \frac{1}{1+\frac{1}{2\sin^2(\theta)} \bar{\gamma}_k|d_{\min}|^2} \dd \theta=\\
\sum \limits_{k=0}^{R} c_k \bar{\gamma}_k \int \limits_{0}^{\frac{\pi}{2}} \frac{1}{1+\frac{1}{2\sin^2(\theta)} \bar{\gamma}_k |d_{\min}|^2} \dd \theta=\\
\frac{1}{2} \sum \limits_{k=0}^{R}  \frac{\bar{\gamma}_k^R}{\prod \limits_{\substack{j=0 \\j\neq k}}^{R}(\bar{\gamma}_k-\bar{\gamma}_j)}
\left\lbrace
1-\sqrt{\frac{\bar{\gamma}_k |d_{\mathrm{min}}|^2}{2+\bar{\gamma}_k |d_{\mathrm{min}}|^2}}
\right\rbrace
\end{multline}

Now, for the second case that $\bar{\gamma}_0 = \bar{\gamma}_i=\bar{\gamma}, \forall \; i>0$, again using the partial fraction technique gives
\begin{multline}
\label{eq:par-frac-case2}
\frac{1}{ 1+\frac{1}{2\sin^2(\theta)}\bar{\gamma}_0|d_{\mathrm{min}}|^2 }
\prod \limits_{i=1}^{R} \frac{1}{ 1+\frac{1}{2\sin^2(\theta)}\bar{\gamma}_i|d_{\mathrm{min}}|^2 }=\\
\left(
\frac{1}{ 1+\frac{1}{2\sin^2(\theta)}\bar{\gamma}|d_{\mathrm{min}}|^2 }
\right)^{R+1}
\end{multline}

Hence, using the integral techniques in \cite{integral-tables}, one obtains
\begin{multline}
\label{eq:err-floor-proof2}
\frac{1}{\pi} \int \limits_{0}^{\frac{\pi}{2}}
\left(
\frac{1}{ 1+\frac{1}{2\sin^2(\theta)}\bar{\gamma}|d_{\mathrm{min}}|^2 }
\right)^{R+1}\dd \theta=\\
\frac{1}{2} \left\lbrace
1- \sqrt{\frac{\bar{\gamma}|d_{\mathrm{min}}|^2}{\bar{\gamma}|d_{\mathrm{min}}|^2+2}} \sum \limits_{l=0}^{R} \binom{2l}{l} \left( \frac{1}{4+2\bar{\gamma}|d_{\mathrm{min}}|^2} \right)^l
\right\rbrace
\end{multline}

For the last case $\; \bar{\gamma}_0\neq \bar{\gamma}_i=\bar{\gamma}, \forall \; i>0$, one has
\begin{multline}
\label{eq:par-frac-case3}
\frac{1}{ 1+\frac{1}{2\sin^2(\theta)}\bar{\gamma}_0|d_{\mathrm{min}}|^2 }
\prod \limits_{i=1}^{R} \frac{1}{ 1+\frac{1}{2\sin^2(\theta)}\bar{\gamma}_i|d_{\mathrm{min}}|^2 }=\\
\frac{1}{ 1+\frac{1}{2\sin^2(\theta)}\bar{\gamma}_0|d_{\mathrm{min}}|^2 }
\left(
\frac{1}{ 1+\frac{1}{2\sin^2(\theta)}\bar{\gamma}|d_{\mathrm{min}}|^2 }
\right)^{R}=\\
\frac{b_0}{1+\frac{1}{2\sin^2(\theta)} \bar{\gamma}_0 |d_{\min}|^2}+
\sum \limits_{k=1}^{R} \frac{b_k}{\left(1+\frac{1}{2\sin^2(\theta)}\bar{\gamma} |d_{\min}|^2\right)^k}
\end{multline}
where $b_0=\left(\frac{\bar{\gamma}_0}{\bar{\gamma}_0-\bar{\gamma}}\right)^R$ and $b_k=\frac{-\bar{\gamma}_0^{R-k}\bar{\gamma}}{(\bar{\gamma}_0-\bar{\gamma})^{R-k+1}}$. Then taking the integration from \eqref{eq:par-frac-case3} gives the error floor expression in \eqref{eq:PEP-floor3}.

\balance
\bibliographystyle{IEEEbib}
%\bibliography{h:/latex/references}
\bibliography{ref/references}

\vspace*{1cm}

\end{document}